\documentclass[%
    11pt,
    DIV=11,
    BCOR=0mm,
    twoside=semi,
    headings=small
]{scrartcl}

\usepackage{scrpage2}

\addtokomafont{title}{\normalfont \bfseries}
\addtokomafont{section}{\normalfont \bfseries}
\addtokomafont{subsection}{\normalfont \bfseries}
\addtokomafont{subsubsection}{\normalfont \bfseries}
\addtokomafont{paragraph}{\normalfont \bfseries}
\newcommand{\quoteparagraph}[1]{\noindent{\normalsize \bfseries #1}\hspace{1em}}

\usepackage[utf8]{inputenc}           
\usepackage{amsmath, amsthm, amssymb} 
\usepackage[english]{babel}           
\usepackage{bibgerm}                  
\usepackage{graphicx}                 
\usepackage{hyperref}                 
\usepackage{booktabs}
\usepackage{siunitx}
\usepackage{color}
\usepackage[table]{xcolor}
\usepackage{rotating}

\newcommand{\R}{\ensuremath{\mathbb{R}}}

\newcommand{\N}{\ensuremath{\mathbb{N}}}

\newtheorem{definition}{Definition}
\newtheorem{remark}{Remark}

\usepackage{bookmark}
\hypersetup{
  pdftitle    = {Lead-Lag Relationship using a Stop-and-Reverse-MinMax Process},
  pdfauthor   = {Stanislaus Maier-Paape and Andreas Platen},
  pdfkeywords = {lead-lag relationship, intermarket analysis, local extrema, empirical distribution},
  pdfborder={0 0 0.5}
}

\author{
  \normalsize \textsc{Stanislaus Maier-Paape}\\[-0.2em]
    \small \textit{Institut f\"ur Mathematik, RWTH Aachen,}\\[-0.5em]
    \small \textit{Templergraben 55, D-52052 Aachen, Germany}\\[-0.5em]
    \small \href{mailto:maier@instmath.rwth-aachen.de}{maier@instmath.rwth-aachen.de}\\
  \\[-0.75em]
  \normalsize \textsc{Andreas Platen}\\[-0.2em]
    \small \textit{Institut f\"ur Mathematik, RWTH Aachen,}\\[-0.5em]
    \small \textit{Templergraben 55, D-52052 Aachen, Germany}\\[-0.5em]
    \small \href{mailto:platen@instmath.rwth-aachen.de}{platen@instmath.rwth-aachen.de}
}
\date{
  \vspace{0.25em}
  \normalsize\today
  \vspace{-1cm}
}
\title{
  \vspace{-2cm}
  \Large Lead-Lag Relationship using a Stop-and-Reverse-MinMax Process
}

\clearscrheadfoot
\lehead{\pagemark}                   
\rohead{\pagemark}                   
\begin{document}

\maketitle

\begin{quote}
  \small
  \quoteparagraph{Abstract}
  The intermarket analysis, in particular the lead-lag relationship, plays an important role within financial markets. Therefore a mathematical approach to be able to find interrelations between the price development of two different financial underlyings is developed in this paper. Computing the differences of the relative positions of relevant local extrema of two charts, i.e., the local phase shifts of these underlyings, gives us an empirical distribution on the unit circle. With the aid of directional statistics such angular distributions are studied for many pairs of markets. It is shown that there are several very strongly correlated underlyings in the field of foreign exchange, commodities and indexes. In some cases one of the two underlyings is significantly ahead with respect to the relevant local extrema, i.e., there is a phase shift unequal to zero between these two underlyings.
  \\

  \quoteparagraph{Keywords} lead-lag relationship, intermarket analysis, local extrema, empirical distribution
\end{quote}



\section{Introduction}

It is well-known that financial markets can be strongly correlated in such a way that their market values show a similar behavior. Knowing the exact connection between two markets would be very helpful for risk-averse investment strategies. In case that two markets are perfectly correlated it would make no difference to invest in either one of them or both together. One simply cannot diversify the risk on both markets. In case it is known that one market leads the other, one is able to use the leading market as an indicator to predict the price development of the other market. Knowing this connection between the two markets can be useful to improve the investment strategy. Therefore we develop a method for quantizing the interrelation of two markets from a different point of view: We want to be able to identify a possible phase shift between two markets if they are correlated.

This subject has been approached in a variety of articles. One approach is to decompose the time series of two markets on a scale-by-scale basis into components with different frequencies using wavelets. The lead-lag relationship is studied by comparing the components of one selected level of the wavelet transformation for two markets, see e.g.
\cite{Dajcman2013,Fiedor2014,IK2006,KI2005,RL1998a,RL1998}.
More on wavelet methods in finance can be found in the book of Gen\c{c}ay, Sel\c{c}uk and Whitcher~\cite{GSW2001}.

Other methods working with correlation, auto-correlation and similar quantities can be found e.g. in~\cite{Chan1993,JD1998,JN1997,EK2011,GT2011,Iwaisako2007,SW1990}.
Didier, Love and Per\'ia~\cite{DLP2012} studied the comovements during 2007--2008 crisis. A different but related topic is the lead-lag relationship between news, e.g. on Twitter, and stock prices, see e.g. \cite{BMZ2011,MCB2011}.

For the intermarket analysis from a point of view of the technical analysis see e.g. the book of Murphy~\cite{Murphy2004} and also of Ruggiero~\cite{Ruggiero1997}.

However, to the best of the author's knowledge, the approaches found in the literature do not follow a geometric approach, e.g. they do not take local extreme values of the time series into account.
Decomposing the time series using wavelets permits to write the time series as the sum of wavelike components with different frequency spectrum. Using these components for comparison of different markets will therefore compare only parts of the original time series. The problem is that these components can be hidden in the original time series such that a possible lag observed between the components of the same level does not necessarily mean that this lag can be observed in the time series itself, e.g. by comparing reversal points. Therefore it is not clear how to interpret the results with regards to an application.

Since we want to be able to receive results giving us an observable lead-lag relationship of two time series, we prefer a geometrical approach. For this reason we need significant points to be able to uniquely identify a lead or lag if any. Very important situations are reversal points and thus the points in time of relevant local extreme values which represents the moment of reversal. A possible lead or lag can then directly be seen by comparing the local extrema of both charts. Such an ansatz could be used for trading these financial products and offers a deep insight into the lead-lag relationship between two markets because an empirical distribution over all local phase shifts can be identified. Additionally the results are not hidden in just one single value like cross-correlation.

The paper is organized as follows: The search for the relevant local extreme values is far from being unique. Therefore we discuss in Section~\ref{sec:2} the approach to find these extreme values for a given pair of markets which we want to compare. Using these values we can compute local phase shifts of both markets which gives us a corresponding empirical distribution. To analyze the results we introduce the directional statistics in Section~\ref{sec:3}. Now we can apply our approach to historical data, e.g. for foreign exchange, commodities and indexes, which we do in Section~\ref{sec:4}. In Section~\ref{sec:5} we give some conclusions.


\section{Method for intermarket analysis}\label{sec:2}

Suppose we want to compare two financial underlyings namely market $A$ and market $B$ for lead and lag. First we take one chart for each underlying with the same bar size, e.g. a \SI{60}{\minute} chart, depending on our interest. Now we want to decide whether these two charts are correlated and show lead or lag. Of course if both underlyings are fully uncorrelated we cannot compare them. Therefore let us assume that there is a connection between these two charts.

Since we prefer a geometrical ansatz we need the points in time of relevant local extreme values. If each maximum occurs for both charts at the exact same time and the same holds true for the minimal values we can say that both underlyings run perfectly synchronous. If the maximum of chart $B$ occurs shortly after the maximum of chart $A$, we observe that market $B$ has a lag compared to market $A$.

Such a comparison could easily be done by hand in a very intuitive way. If one compares two markets and gets a feeling for lead-lag relationship, e.g. assume market $A$ leads $B$, one directly benefits from this knowledge because right after a reversal point in market $A$ would most likely occur a reversal point in market $B$. This can be very useful for several strategies (for position entries and also for exits).

Of course doing an extensive study by hand would be very time consuming and not objective. For an automatic approach we first need an appropriate method to identify local extrema for both time series. The MinMax algorithm introduced by Maier-Paape~\cite{Maier-Paape2013} is a method which yields such a series of alternating relevant local extrema (called MinMax process) and will therefore be used in the following. This method uses a so called SAR (stop and reverse) process to identify up and down movements. If an up movement is detected the MinMax algorithm searches for a maximum and fixes this local maximal value if the movement phase reverses to a down movement. Minimal values are searched during down movement phases. The underlying SAR process could be the MACD (moving average convergence/divergence) indicator of~\cite{Appel2005} which, simplified speaking, indicates an up movement if the MACD series lies above its signal line and a down movement when its vice versa. See \cite{Maier-Paape2013} for the details.

The SAR process controls the frequency of detected local extreme values and, in general, is controlled by some parameters (default for MACD are 12, 26 and 9). In this paper we will always use the MACD as SAR process. Instead of adjusting several parameters separately we use just a common factor, called timescale, that scales the three default parameters at the same time. Increasing the timescale leads to less extreme values while decreasing timescale leads to more extreme values, i.e. a finer resolution.

Note that the MACD series can oscillate quickly around the signal line which leads to many small and insignificant local extreme values. To avoid this problem we require for a change of the direction of the SAR process that the distance of MACD and its signal line needs to exceed some minimal threshold of $\delta = 0.3 \cdot \mbox{ATR} (100)$, where ATR means the average true range, see \cite[Subsection 2.1]{Maier-Paape2013} for the details.

From now on we use this MinMax algorithm because this is a very flexible tool to identify local extreme values. As far as we know this method is the only one which identifies local extreme values exactly and is continuously adjustable. Since a financial time series always has some noise there is no unique objective choice for relevant local extrema of a financial time series. Therefore this process needs to be parameter dependent to adjust the resolution of the minima and maxima.

One question is how to choose the ``right'' parameter. This will be discussed at the end of this section.
For the moment let us assume we already found ``good'' parameters for market $A$. The MinMax process then yields consecutive minima and maxima denoted by $(t_i,X_i)_{i=1,...,N}$ with points in time $t_1\leq ...\leq t_N$ and consecutive price values $X_i$. To be able to compare these points, we measure the time in seconds since 1st January 1970. For this wavelike time series we can compute the mean wavelength by
\begin{align}\label{eq:mean_wavelength}
  \lambda := \frac{1}{N-1}\sum_{i=1}^{N-1} 2(t_{i+1}-t_{i}) = 2\frac{t_{N}-t_{1}}{N-1}.
\end{align}
Note that $\lambda$ depends on the parameters used in the MinMax algorithm since the minima and maxima depend on the used parameters.

Fixing these parameters for the second market gives us the extreme values $(\tilde{t}_i,\tilde{X}_i)_{i=1,...,\tilde{N}}$ with mean wavelength $\tilde{\lambda}$. Of course it makes no sense to compare both markets using these extreme values for very different wavelengths $\lambda$ and $\tilde{\lambda}$. Therefore we fit the parameters of the MinMax process for market $B$ so that $\tilde{\lambda}=\lambda$ holds true.

\begin{remark}
  Note that in general we can not expect a constant but only a time dependent wavelength which can vary a lot, see Figure \ref{fig:timedependet_wavelength}, where the moving average of wavelengths over $N-1=49$ cycles is shown, i.e. $\frac{1}{49}\sum_{i=s-49}^{s-1} 2(t_{i+1}-t_{i})$ where $s$ is the current time index. Therefore matching the mean wavelength for both markets means just matching the level of refinement and not the position of the extreme values themselves.

  \begin{figure}
    \centering
    \includegraphics{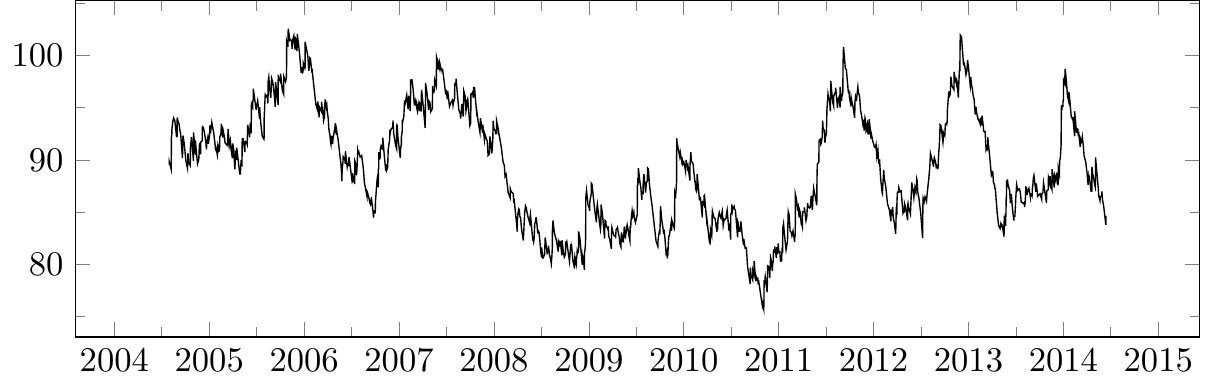}
    \caption{Moving average of wavelengths over $N-1=49$ cycles for S\&P 500 on a \SI{60}{\minute} chart.}
    \label{fig:timedependet_wavelength}
  \end{figure}
\end{remark}

\begin{figure}
  \centering
  \includegraphics{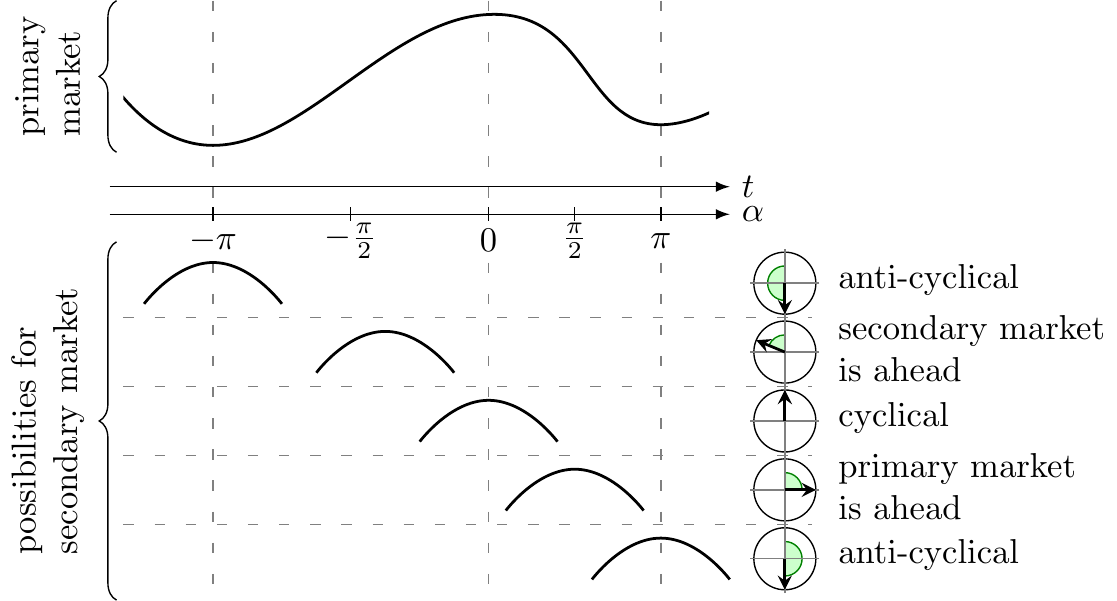}
  \caption{Computation of $\alpha$ in \eqref{eq:alpha}.}
  \label{fig:omega}
\end{figure}

Since we are interested in the lead-lag relationship between market $A$ and $B$ we only need to find the relationship of points in time of the extrema by finding the relative positions of $(\tilde{t}_i)_{i=1,...,\tilde{N}}$ within $(t_j)_{j=1,...,N}$. In this case we call market $A$ the primary market and market $B$ the secondary market. The overall procedure is as follows:

\begin{enumerate}
  \item Fix the desired mean wavelength $\lambda^*>0$.
  \item Find all local extreme values $(t_i,X_i)_{i=1,...,N}$ and  $(\tilde{t}_j,\tilde{X}_j)_{j=1,...,\tilde{N}}$ for the primary and the secondary market, respectively, such that the mean wavelengths~\eqref{eq:mean_wavelength} for both markets on the full data base matches $\lambda^*$, i.e. that we have
    $2\frac{t_{N}-t_{1}}{N-1} \approx \lambda^* \approx 2\frac{\tilde{t}_{\tilde{N}}-\tilde{t}_{1}}{\tilde{N}-1}$.
  \item Find $j_1,j_2\in\{1,...,\tilde{N}\}$ such that $\tilde{t}_{j_1}=\min\{\tilde{t}_j\,:\,\tilde{t}_j\geq t_1\}$ and $\tilde{t}_{j_2}=\max\{\tilde{t}_j\,:\,\tilde{t}_j< t_{N}\}$.
  \newline
  For each $j\in\{j_1,...,j_2\}$ do the following:
  \begin{enumerate}
    \item Find $i\in\{1,...,N-1\}$ such that $t_i\leq \tilde{t}_j <t_{i+1}$.
    \item Define the phase shift of extreme value $(\tilde{t}_j,\tilde{X}_j)$ regarding the extreme values $(t_i,X_i)$ and $(t_{i+1},X_{i+1})$. Here we use the linear relative distance between the corresponding extrema values measured as an angle. We set
    \begin{align}\label{eq:alpha}
      \alpha_j^{\lambda^*} &:= \frac{\tilde{t}_j - s_j}
                        {t_{i+1} - t_i} \cdot\pi\in[-\pi,\pi),
    \end{align}
    where
    \begin{align}
      s_j :=\begin{cases}
                t_i,&\text{if } X_i \text{ and } \tilde{X}_j \text{ are both maxima or both minima},\\
                t_{i+1},&\text{if } X_{i+1} \text{ and } \tilde{X}_j \text{ are both maxima or both minima}.
              \end{cases}
    \end{align}
    Figure~\ref{fig:omega} shows some examples for the position of a maximum of the secondary market relative to some extreme values of the primary market.
\end{enumerate}
  \item We end up with the empirical circular distribution $(\alpha_j^{\lambda^*})_{j=j_1,...,j_2}\subset[-\pi,\pi)$ depending on the mean wavelength $\lambda^*$.
\end{enumerate}

Negative $\alpha$ resemble a front-running (lead) of the secondary market, positive $\alpha$ resemble a time lag of the secondary market. The result can be interpreted on the unit sphere $S^1=\{(\sin\alpha,\cos\alpha)\in\R^2\,:\,\alpha\in[-\pi,\pi)\}$ and gives us all observations of local phase shifts between two markets.

\begin{remark}
  This approach is independent of the openings of the stock exchange for market $A$ and market $B$. Since we measure the points in time $t_i$ and $\tilde{t}_j$ in seconds since 1st January 1970 we just put these values into \eqref{eq:alpha} and the machinery works straight forward.
\end{remark}

\begin{remark}\label{rem:parameter_choice}
  The above method has only one parameter, namely the mean wavelength $\lambda^*$, see step 1. Therefore we can compute different distributions for different wavelengths. It turns out that the results in most cases do not depend on the wavelength. Therefore we compute $(\alpha_i^{\lambda})_{i=1,...,n(\lambda)}$ for many values of the mean wavelength $\lambda$. For each $\lambda$ we can generate a histogram or rather a bar plot and at the end we can compute the average of all bars including standard deviation.
\end{remark}

\begin{remark}\label{rem:extrema_confirmed}
  Note that the extreme values cannot be determined in real time. There is always at least a small time lag. Therefore we can also identify such an empirical distribution if we use the point in time when the extreme value is confirmed by the MinMax algorithm instead of the point in time of the extreme value itself.
\end{remark}


\section{Directional statistics}\label{sec:3}

Since we work with circular distributions, the mean and variance must be computed in an appropriate way, see e.g. \cite{Fisher1996,MJ1999}. This can be used to identify a possible phase shift. We introduce the basic statistical quantities in Subsection~\ref{sec:3.1}. For a deeper analysis we list some interesting statistical tests in Subsection~\ref{sec:3.2} and give an approximation of the lead or lag in Subsection~\ref{sec:3.3}.

\subsection{Basic quantities}\label{sec:3.1}

Now we will discuss how to calculate estimators, e.g. for the mean angular direction. Details on computations for a general distribution with a $2\pi$ periodic probability density function $f$ can be found in \cite[Section 3.2]{Fisher1996}.

The first step is to identify the angles by vectors on the unit sphere $S^1$. Let $(\alpha_j)_{j=1,...,N}\subset[-\pi,\pi)$ be the outcomes of a discrete distribution for the phase shift of two markets of interest. We can identify each angle $\alpha_j$ with a point on the unit sphere
\begin{align*}
  \mathbf{r}_j &:=\left(\begin{matrix}\sin\alpha_j \\ \cos\alpha_j\end{matrix}\right) \in S^1
\end{align*}
for $j=1,...,N$. In this two-dimensional space we can compute the \textit{mean resultant vector} which is defined by
\begin{align*}
  \hat{\mathbf{r}} &
      := \frac{1}{N}\sum_{j=1}^N \mathbf{r}_j.
\end{align*}
Note that for the length of $\hat{\mathbf{r}}$ we have $\|\hat{\mathbf{r}}\|_2\leq 1$ because it is a convex combination of vectors in $S^1$. If $\hat{\mathbf{r}}\neq 0$ choose the \textit{mean angular direction} $\hat{\alpha}\in[-\pi,\pi)$ such that
\begin{align}\label{eq:NR}
  \left(\begin{matrix}\sin \hat{\alpha} \\ \cos \hat{\alpha}\end{matrix}\right) &
      = \frac{1}{\|\hat{\mathbf{r}}\|_2}\hat{\mathbf{r}}.
\end{align}

Of course $\hat{\mathbf{r}}$ could be zero and thus no unique mean angular direction would exist. This is the case, e.g., if the angles are uniformly distributed all around $S^1$. If this is the case for the phase shifts between two markets then there is no connection between them and the analysis of the results would already be finished. Since we are interested in at least slightly correlated markets we do not expect this behavior.

Nevertheless even in the case where $\|\hat{\mathbf{r}}\|_2>0$, the length of $\hat{\mathbf{r}}$ could be small. This happens if the outcomes of the distribution have a large variance. In contrast a length of $\hat{\mathbf{r}}$ near $1$ indicates a small variance and a high concentration of the outcomes near to its mean angular direction. Therefore we need to consider the \textit{circular variance} (cf. \cite[Section~2.3.1, Equation (2.11)]{Fisher1996}) which can be defined by
\begin{align*}
  \hat{S}&:=1-\|\hat{\mathbf{r}}\|_2 \in [0,1].
\end{align*}
To be able to also measure the skewness and the peakedness we define the \textit{circular skewness} by
\begin{align*}
  \hat{b}&:=\frac{1}{N}\sum_{j=1}^N \sin(2(\alpha_j-\hat{\alpha})) \in [-1,1]
\end{align*}
and the \textit{circular kurtosis} by
\begin{align*}
  \hat{k}&:=\frac{1}{N}\sum_{j=1}^N \cos(2(\alpha_j-\hat{\alpha})) \in [-1,1].
\end{align*}

\begin{figure}
  \centering
  \includegraphics[width=15.2cm]{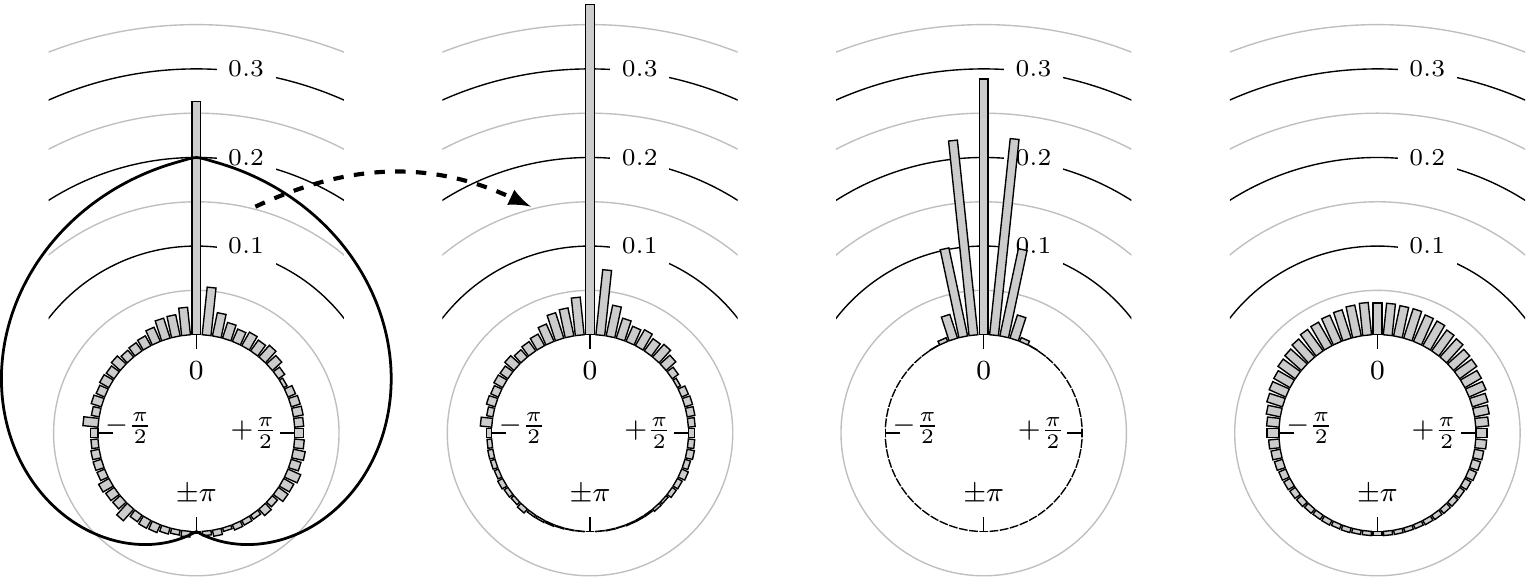}
  \caption{First: Example of a possible distribution of phase shifts and a hat function on $S^1$; Second: Corresponding weighted version of the example distribution from first plot; Third: Plot of the probability density functions of von Mises distributions mean location parameter $\mu=0$ and concentration parameters $\kappa=50$; Fourth: Same as third but with $\kappa=1$.}
  \label{fig:von_mises}
\end{figure}

Since we are interested in the possible lead or lag between two markets we want to reduce the influence of outliers which are far away from the mean angular direction. For this reason we use a hat function on $S^1$ to weight the empirical distribution with the hat near the position of the highest peak of the distribution. Then all reasonable data near the peak get high weights and thus more influence in our statistics, while less important data, i.e. the outliers, obtain small weights.
We expect that the peaks of the distributions are near zero up to some lead or lag, i.e. the two markets are positive correlated. Therefore we use the hat function which has its hat (maximum) at zero and is zero (minimum) at $\pm\pi$. The first two plots of Figure~\ref{fig:von_mises} show an example for an observed distribution and its weighted counterpart, respectively. From the weighted distribution we can compute the \textit{weighted mean angular direction} $\hat{\alpha}^{(w)}$ as in \eqref{eq:NR}.

\subsection{Statistical tests}\label{sec:3.2}

Most of the statistical tests require an underlying von Mises distribution, see e.g. \cite[Section 3.3.6]{Fisher1996}, which is often used as an analogon to normal distribution on the unit sphere. The distribution we get for our application is not exactly a von Mises distribution but has a similar shape, see Figure~\ref{fig:von_mises}. In this figure the distribution of phase shifts has a similar shape to two superposed von Mises distribution, one with a large and one with a small concentration parameter $\kappa$. Thus it is possible, that the phase shifts correspond to a von Mises distribution plus noise, e.g. white noise. Nevertheless we use the following statistical tests in order to be able to classify the results even if they are designed for von Mises distributions.

Since we do not know the underlying distribution for the phase shifts we only get some realizations. Computing the quantities in Section~\ref{sec:3.1} using the formulas by putting in our observations will give us the estimators which will be denoted by $\hat{\alpha}$, $\hat{\alpha}^{(w)}$, $\hat{S}$, $\hat{b}$ and $\hat{k}$, respectively.

Next we want to verify the quality of our mean angular direction.
Therefore we compute the $(1-\delta)\si{\percent}$-confidence intervals for the population mean, such that $L_1:=\hat{\alpha}-d$ and $L_2:=\hat{\alpha}+d$ are the lower and upper confidence limits of the mean angular direction, respectively, see \cite[Section 26.7]{Zar2010}. For the weighted mean $\hat{\alpha}^{(w)}$ we denote the confidence interval by $d^{(w)}$. We always use $\delta=\SI{5}{\percent}$.

To test for zero mean which would imply that there is no lead or lag relationship we can perform the one sample test for mean angle, which is similar to the one sample $t$-test on a linear scale. Let $\alpha_0\in[-\pi,\pi)$ be the mean angular direction for which we want to test and $\bar{\alpha}$ the mean angular direction of the underlying (unknown) distribution. We test for
\begin{align*}
  H_0:\,\,&\bar{\alpha}=\alpha_0,\\
  H_1:\,\,&\bar{\alpha}\neq\alpha_0
\end{align*}
by checking whether $\alpha_0\in[L_1,L_2]$ using our estimator $\hat{\alpha}$ and its \SI{95}{\percent} confidence interval, see \cite[Section 27.1 (c)]{Zar2010}. In our case we will set $\alpha_0=0$. The result of this test is then given by
\begin{align*}
  h_m &:= \begin{cases}
            0,&\text{if }H_0\text{ can not be rejected, i.e. } \alpha_0=0\in [L_1,L_2],\\
            1,&\text{otherwise}.
          \end{cases}
\end{align*}

In Remark~\ref{rem:parameter_choice} we noted that we will generate empirical distributions for different mean wavelengths, say $n\in\N$ different values. To compare all these distributions for the same pair of markets we can use the one-factor ANOVA or Watson-Williams test (multi-sample test). It assesses the question whether the mean directions of two or more groups are identical or not, i.e. it tests for
\begin{align*}
  H_0:\,\,&\text{All of $n$ groups share a common mean direction, i.e., }\bar{\alpha}^{(1)}=...=\bar{\alpha}^{(n)}.\\
  H_1:\,\,&\text{Not all groups have a common mean direction,}
\end{align*}
see \cite[Section 27.4 (b)]{Zar2010}.
The output of this test is a $p$-value, i.e. the probability of getting results which are at least as extreme as our observation assuming the null hypothesis is true. Thus a large $p$-value indicates that the null hypothesis holds true. We denote this value by $p_{ww}\in [0,1]$.

\subsection{Lead or lag}\label{sec:3.3}

Using the mean angular direction $\hat{\alpha}$ and its confidence interval we can roughly approximate the lead or lag. Assume we have a mean wavelength of 100 candles on a \SI{60}{\minute} chart. The mean wavelength would then be approximately $\SI{60}{\minute}\cdot 100=\SI{6000}{\minute}$. This value equates $2\pi$. Thus the mean of the lead or lag $\ell$ can be approximated by
\begin{align*}
  \ell \approx \frac{\hat{\alpha}}{2\pi}\cdot \SI{6000}{\minute}
\end{align*}
and the corresponding confidence interval is approximated by $[\ell-d_{\ell},\ell+d_{\ell}]$ where
\begin{align*}
  d_{\ell} \approx \frac{d}{2\pi}\cdot \SI{6000}{\minute}.
\end{align*}
Analogously we can compute the lead or lag using the weighted mean angular direction which we denote by $\ell^{(w)}$ and $d_{\ell}^{(w)}$, respectively, i.e. $\ell^{(w)}\approx\frac{\hat{\alpha}^{(w)}}{2\pi}\cdot \SI{6000}{\minute}$ and $d_{\ell}^{(w)}\approx \frac{d^{(w)}}{2\pi}\cdot \SI{6000}{\minute}$. Note that a positive value for $\ell$ and $\ell^{(w)}$ means that the primary market leads the secondary and vice versa for a negative value.

To answer the question which market is ahead, if any, we make the following definition.
\begin{definition}\label{def:lead_lag}
  For positive correlated markets, i.e. $|\hat{\alpha}^{(w)}|\leq\frac{\pi}{2}$, we say \emph{one market leads the other} if $\hat{\alpha}^{(w)}$ is significantly different from zero, i.e.
  \begin{align*}
    \text{if }\,\, \hat{\alpha}^{(w)}-d^{(w)}>0 &\quad\leadsto\quad \text{primary market leads,}\\
    \text{if }\,\, \hat{\alpha}^{(w)}+d^{(w)}<0 &\quad\leadsto\quad \text{secondary market leads.}
  \end{align*}
\end{definition}


\section{Empirical study}\label{sec:4}

Now we study different markets from commodities to foreign exchanges. In Subsection~\ref{sec:4.1} we explain the setting and give some details on the choice of parameters. The angular histograms and the statistical results are then shown in Subsection~\ref{sec:4.2}.

\subsection{Settings}\label{sec:4.1}

In this paper we focus on the \SI{60}{\minute} chart. The wavelengths we use to adjust the MinMax process for the primary market, see Remark~\ref{rem:parameter_choice}, are of size of 30 candles up to 180 candles. For the Watson-Williams test, see Section~\ref{sec:3.2}, this leads to $n=151$ groups.

Since we have given the wavelength in number of candles we proceed as follows to ``synchronize'' the markets:
\begin{enumerate}
  \item Choose the desired mean wavelength $\lambda^*_{\text{candles}}\in\{30,31,...,180\}$ in number of candles.
  \item Adjust the parameter for the MinMax process on the primary market, such that the wavelength of the primary market in number of candles, ignoring the time when the stock exchange is closed, matches $\lambda^*_{\text{candles}}$.
  \item Calculate the corresponding wavelength $\lambda^*$ in seconds for the primary market, this time considering the time when stock exchange is closed.
  \item Adjust the MinMax process on the secondary market, such that the wavelength of the secondary market in seconds matches $\lambda^*$, i.e. perform step 2 from Section~\ref{sec:2}, where the primary market is already fixed.
  \item Proceed with steps 3 and 4 from Section~\ref{sec:2}.
\end{enumerate}

For most computations of the directional statistics the MATLAB library CircStat~\cite{Berens2009} has been used and all angles are measured in radian.

The markets which we examine including the period of time for the available candle data are listed in Table~\ref{tab:markets}. Note that the start date is not the same for all markets. If we examine a combination of markets with different initial dates we use the smaller period of time for both markets.

\begin{table}
\centering
  { \footnotesize
    \begin{tabular}{llr}
      \toprule
      market              & underlying & initial date\\
      \midrule
      Eurex DAX           & DAX Futures                 & December  10, 2003\\
      Eurex BUND          & Euro-Bund Futures           & December  10, 2003\\
      Eurex DJEST50       & Euro STOXX 50 Index Futures & December  10, 2003\\
      CME MINI S\&P       & E-mini S\&P 500 Futures     & December  12, 2003\\
      CME MINI NSDQ       & E-mini NASDAQ 100 Futures   & September 14, 2004\\
      CME CMX GLD         & Gold Futures                & July \phantom{0}7, 2005\\
      CME CMX SIL         & Silver Futures              & July \phantom{0}7, 2005\\
      CME PH CRDE         & Crude Oil Futures           & November  29, 2004\\
      CME PH NG           & Natural Gas (Henry Hub) Physical Futures & November  29, 2004\\
      CME\_CBT 30Y TB     & U.S. Treasury Bond Futures  & October   18, 2004\\
      ICE\_NYBOT MNRUS2K  & Russell 2000 Index Futures  & September 20, 2004\\
      Forex EUR-USD       & EUR-USD                     & July      17, 2009\\
      Forex JPY-USD       & JPY-USD                     & July      17, 2009\\
      Forex GBP-USD       & GBP-USD                     & July      17, 2009\\
      Forex CHF-USD       & CHF-USD                     & July      17, 2009\\
      \bottomrule
    \end{tabular}
  }
  \caption{Examined markets and the period of time of the used candle data of the \SI{60}{\minute} chart (terminal date is always May 15, 2014). All historical data are from FIDES.}
  \label{tab:markets}
\end{table}

\subsection{Results}\label{sec:4.2}

Now we look at the results for several futures, indexes and foreign exchanges. The statistical quantities for the phase shift of the extreme values are shown in Table~\ref{tab:60min} and for the points in time of the confirmation of the extreme values in Table~\ref{tab:60min_confirmed}. The corresponding empirical distributions are given, according to the following remark, by Figures~\ref{fig:res1} to \ref{fig:res17}.

\begin{remark}\label{rem:figures}
  (Notes on figures)\\
  The label of each of the following figures states ``$A$ versus $B$'' and each figure shows the following four distributions (in same order):
  \begin{enumerate}
    \item Time of extrema: $A$ as primary and $B$ as secondary market.
    \item Time of extrema: $B$ as primary and $A$ as secondary market.
    \item Time of extrema confirmed (see Remark~\ref{rem:extrema_confirmed}): $A$ as primary and $B$ as secondary market.
    \item Time of extrema confirmed (see Remark~\ref{rem:extrema_confirmed}): $B$ as primary and $A$ as secondary market.
  \end{enumerate}
  All plots also contains the mean angular direction and the mean angular direction of the weighted distribution (weighted with the hat function, see Figure~\ref{fig:von_mises}). These directions are the green and red line inside the circle, respectively.

  Additionally each bin of the histograms contains information of the single distributions for each wavelength: It shows the largest value of this bin occurred within the $151$ single distributions, the smallest value and the bin value of the combined distribution plus and minus the standard deviation.
\end{remark}

\begin{sidewaystable}
  {\footnotesize
   \def\ROWCOLOR{black!10!white}
   \newcommand{\mc}[1]{\multicolumn{1}{c}{#1}}
   \newcommand{\bb}[1]{#1$^*$}
   \begin{tabular}{llrrrrrrrc}
   \toprule
   prime & sec & \mc{$\hat{\alpha}\pm d$} & \mc{$\hat{\alpha}^{(w)} \pm d^{(w)}$} & $\ell^{(w)}\pm d_{\ell}^{(w)}$/\si{\minute} & \mc{$\hat{S}$} & \mc{$\hat{b}$} & \mc{$\hat{k}$} & \mc{$p_{ww}$} & \mc{$h_m$}\\
   \midrule
   \rowcolor{\ROWCOLOR}
       \bb{Eurex DAX} &      CME MINI S\&P & $  0.002 \pm   0.008$ & $  0.012 \pm   0.003$ & $   11.833 \pm   2.674$ &   0.553 &   0.028 &   0.349 &   1.000 & 0\\
   \rowcolor{\ROWCOLOR}
        CME MINI S\&P &     \bb{Eurex DAX} & $  0.035 \pm   0.006$ & $ -0.005 \pm   0.002$ & $   -4.809 \pm   2.001$ &   0.522 &  -0.084 &   0.426 &   1.000 & 1\\
        Forex EUR-USD & \bb{Forex JPY-USD} & $ -0.286 \pm   0.081$ & $ -0.044 \pm   0.006$ & $  -41.693 \pm   5.571$ &   0.948 &   0.071 &   0.161 &   0.000 & 1\\
   \bb{Forex JPY-USD} &      Forex EUR-USD & $  0.291 \pm   0.072$ & $  0.028 \pm   0.006$ & $   26.967 \pm   5.572$ &   0.941 &  -0.098 &   0.142 &   0.000 & 1\\
   \rowcolor{\ROWCOLOR}
        Forex EUR-USD & \bb{Forex GBP-USD} & $  0.002 \pm   0.011$ & $ -0.010 \pm   0.003$ & $   -9.381 \pm   3.303$ &   0.617 &  -0.031 &   0.342 &   1.000 & 0\\
   \rowcolor{\ROWCOLOR}
   \bb{Forex GBP-USD} &      Forex EUR-USD & $  0.025 \pm   0.011$ & $  0.007 \pm   0.004$ & $    6.269 \pm   3.370$ &   0.637 &  -0.024 &   0.341 &   1.000 & 1\\
   \bb{Forex EUR-USD} &      Forex CHF-USD & $  0.000 \pm   0.008$ & $  0.005 \pm   0.003$ & $    4.645 \pm   2.672$ &   0.489 &   0.012 &   0.513 &   1.000 & 0\\
        Forex CHF-USD & \bb{Forex EUR-USD} & $ -0.003 \pm   0.008$ & $ -0.013 \pm   0.003$ & $  -12.668 \pm   2.724$ &   0.492 &  -0.036 &   0.501 &   1.000 & 0\\
   \rowcolor{\ROWCOLOR}
           Eurex BUND & \bb{CME\_CBT 30Y TB} & $ -0.036 \pm   0.010$ & $ -0.007 \pm   0.003$ & $   -6.534 \pm   3.098$ &   0.613 &   0.038 &   0.306 &   1.000 & 1\\
   \rowcolor{\ROWCOLOR}
   \bb{CME\_CBT 30Y TB} &         Eurex BUND & $  0.078 \pm   0.008$ & $  0.012 \pm   0.003$ & $   11.759 \pm   2.565$ &   0.612 &  -0.099 &   0.331 &   1.000 & 1\\
        CME MINI S\&P & \bb{CME MINI NSDQ} & $ -0.022 \pm   0.005$ & $ -0.009 \pm   0.002$ & $   -8.670 \pm   1.749$ &   0.415 &   0.029 &   0.578 &   1.000 & 1\\
   \bb{CME MINI NSDQ} &      CME MINI S\&P & $  0.035 \pm   0.005$ & $  0.013 \pm   0.002$ & $   12.253 \pm   1.746$ &   0.409 &  -0.050 &   0.574 &   1.000 & 1\\
   \rowcolor{\ROWCOLOR}
   \bb{ICE\_NYBOT MNRUS2K} &      CME MINI S\&P & $  0.019 \pm   0.005$ & $  0.005 \pm   0.002$ & $    5.004 \pm   1.738$ &   0.393 &  -0.034 &   0.596 &   1.000 & 1\\
   \rowcolor{\ROWCOLOR}
        CME MINI S\&P & \bb{ICE\_NYBOT MNRUS2K} & $ -0.012 \pm   0.005$ & $ -0.005 \pm   0.002$ & $   -4.714 \pm   1.709$ &   0.402 &   0.017 &   0.604 &   1.000 & 1\\
   ICE\_NYBOT MNRUS2K & \bb{CME MINI NSDQ} & $ -0.014 \pm   0.005$ & $ -0.012 \pm   0.002$ & $  -11.257 \pm   1.885$ &   0.464 &  -0.004 &   0.542 &   1.000 & 1\\
   \bb{CME MINI NSDQ} & ICE\_NYBOT MNRUS2K & $  0.024 \pm   0.005$ & $  0.008 \pm   0.002$ & $    7.952 \pm   1.858$ &   0.466 &  -0.032 &   0.528 &   0.948 & 1\\
   \rowcolor{\ROWCOLOR}
        Eurex DJEST50 &     \bb{Eurex DAX} & $  0.001 \pm   0.005$ & $ -0.002 \pm   0.002$ & $   -2.368 \pm   2.052$ &   0.323 &  -0.012 &   0.638 &   1.000 & 0\\
   \rowcolor{\ROWCOLOR}
            Eurex DAX &      Eurex DJEST50 & $ -0.007 \pm   0.005$ & $ -0.001 \pm   0.002$ & $   -0.999 \pm   2.058$ &   0.318 &   0.016 &   0.620 &   1.000 & 1\\
     \bb{CME CMX GLD} &        CME CMX SIL & $  0.040 \pm   0.006$ & $  0.010 \pm   0.002$ & $    9.576 \pm   2.109$ &   0.498 &  -0.060 &   0.468 &   1.000 & 1\\
     \bb{CME CMX SIL} &        CME CMX GLD & $  0.014 \pm   0.006$ & $  0.010 \pm   0.002$ & $    9.559 \pm   2.086$ &   0.487 &  -0.002 &   0.491 &   1.000 & 1\\
   \rowcolor{\ROWCOLOR}
     \bb{CME CMX GLD} &      Forex EUR-USD & $  0.154 \pm   0.022$ & $  0.030 \pm   0.005$ & $   28.393 \pm   4.488$ &   0.804 &  -0.091 &   0.209 &   0.861 & 1\\
   \rowcolor{\ROWCOLOR}
        Forex EUR-USD &   \bb{CME CMX GLD} & $ -0.116 \pm   0.022$ & $ -0.044 \pm   0.005$ & $  -42.000 \pm   4.334$ &   0.813 &   0.027 &   0.227 &   0.915 & 1\\
     \bb{CME CMX GLD} &      CME MINI S\&P & $  0.067 \pm   0.030$ & $  0.021 \pm   0.004$ & $   19.585 \pm   3.631$ &   0.892 &  -0.021 &   0.248 &   0.000 & 1\\
        CME MINI S\&P &   \bb{CME CMX GLD} & $  0.018 \pm   0.027$ & $ -0.023 \pm   0.004$ & $  -22.367 \pm   3.573$ &   0.883 &  -0.051 &   0.239 &   0.986 & 0\\
   \rowcolor{\ROWCOLOR}
          CME CMX GLD &     \bb{Eurex DAX} & $ -0.031 \pm   0.029$ & $ -0.022 \pm   0.004$ & $  -21.402 \pm   3.696$ &   0.890 &  -0.016 &   0.226 &   0.005 & 1\\
   \rowcolor{\ROWCOLOR}
       \bb{Eurex DAX} &        CME CMX GLD & $ -0.018 \pm   0.039$ & $  0.006 \pm   0.005$ & $    5.552 \pm   4.967$ &   0.897 &   0.020 &   0.165 &   0.000 & 0\\
     \bb{CME CMX GLD} &        CME PH CRDE & $  0.096 \pm   0.016$ & $  0.028 \pm   0.003$ & $   27.069 \pm   3.142$ &   0.803 &  -0.048 &   0.260 &   1.000 & 1\\
          CME PH CRDE &   \bb{CME CMX GLD} & $ -0.051 \pm   0.014$ & $ -0.033 \pm   0.003$ & $  -31.441 \pm   3.153$ &   0.767 &  -0.008 &   0.264 &   0.907 & 1\\
   \rowcolor{\ROWCOLOR}
          CME PH CRDE &     \bb{Eurex DAX} & $ -0.042 \pm   0.018$ & $ -0.035 \pm   0.003$ & $  -33.130 \pm   3.302$ &   0.822 &  -0.024 &   0.242 &   1.000 & 1\\
   \rowcolor{\ROWCOLOR}
       \bb{Eurex DAX} &        CME PH CRDE & $  0.098 \pm   0.024$ & $  0.046 \pm   0.005$ & $   44.008 \pm   4.301$ &   0.842 &   0.001 &   0.190 &   1.000 & 1\\
          CME PH CRDE & \bb{Forex EUR-USD} & $ -0.114 \pm   0.018$ & $ -0.059 \pm   0.005$ & $  -56.518 \pm   4.366$ &   0.759 &   0.008 &   0.221 &   1.000 & 1\\
   \bb{Forex EUR-USD} &        CME PH CRDE & $  0.149 \pm   0.018$ & $  0.064 \pm   0.004$ & $   61.358 \pm   4.162$ &   0.774 &  -0.033 &   0.224 &   1.000 & 1\\
   \rowcolor{\ROWCOLOR}
          CME PH CRDE &     \bb{CME PH NG} & $ -0.062 \pm   0.025$ & $ -0.022 \pm   0.004$ & $  -20.535 \pm   3.643$ &   0.876 &   0.012 &   0.209 &   0.009 & 1\\
   \rowcolor{\ROWCOLOR}
       \bb{CME PH NG} &        CME PH CRDE & $  0.082 \pm   0.025$ & $  0.012 \pm   0.004$ & $   11.090 \pm   3.588$ &   0.872 &  -0.047 &   0.217 &   0.008 & 1\\
   \bottomrule
   \end{tabular}
  }
  \caption{Results on 60\,\textrm{min} chart (time of extrema). $^*$This market leads the other one.}
  \label{tab:60min}
\end{sidewaystable}

\begin{sidewaystable}
  {\footnotesize
   \def\ROWCOLOR{black!10!white}
   \newcommand{\mc}[1]{\multicolumn{1}{c}{#1}}
   \newcommand{\bb}[1]{#1$^*$}
   \begin{tabular}{llrrrrrrrc}
   \toprule
   prime & sec & \mc{$\hat{\alpha}\pm d$} & \mc{$\hat{\alpha}^{(w)} \pm d^{(w)}$} & $\ell^{(w)}\pm d_{\ell}^{(w)}$/\si{\minute} & \mc{$\hat{S}$} & \mc{$\hat{b}$} & \mc{$\hat{k}$} & \mc{$p_{ww}$} & \mc{$h_m$}\\
   \midrule
   \rowcolor{\ROWCOLOR}
            Eurex DAX & \bb{CME MINI S\&P} & $ -0.137 \pm   0.008$ & $ -0.076 \pm   0.003$ & $  -72.647 \pm   2.722$ &   0.548 &   0.045 &   0.325 &   0.000 & 1\\
   \rowcolor{\ROWCOLOR}
   \bb{CME MINI S\&P} &          Eurex DAX & $  0.039 \pm   0.006$ & $  0.008 \pm   0.002$ & $    8.038 \pm   2.041$ &   0.511 &  -0.053 &   0.392 &   0.000 & 1\\
        Forex EUR-USD & \bb{Forex JPY-USD} & $ -0.572 \pm   0.051$ & $ -0.082 \pm   0.006$ & $  -78.478 \pm   5.384$ &   0.917 &   0.166 &   0.080 &   0.000 & 1\\
   \bb{Forex JPY-USD} &      Forex EUR-USD & $  0.494 \pm   0.047$ & $  0.086 \pm   0.006$ & $   81.842 \pm   5.381$ &   0.910 &  -0.143 &   0.116 &   0.000 & 1\\
   \rowcolor{\ROWCOLOR}
        Forex EUR-USD & \bb{Forex GBP-USD} & $ -0.053 \pm   0.011$ & $ -0.029 \pm   0.004$ & $  -27.782 \pm   3.471$ &   0.622 &   0.012 &   0.289 &   0.000 & 1\\
   \rowcolor{\ROWCOLOR}
   \bb{Forex GBP-USD} &      Forex EUR-USD & $  0.053 \pm   0.011$ & $  0.021 \pm   0.004$ & $   19.941 \pm   3.452$ &   0.615 &  -0.032 &   0.290 &   0.000 & 1\\
        Forex EUR-USD & \bb{Forex CHF-USD} & $ -0.048 \pm   0.007$ & $ -0.025 \pm   0.003$ & $  -24.078 \pm   2.669$ &   0.461 &   0.037 &   0.493 &   0.002 & 1\\
   \bb{Forex CHF-USD} &      Forex EUR-USD & $  0.065 \pm   0.007$ & $  0.033 \pm   0.003$ & $   31.507 \pm   2.723$ &   0.463 &  -0.049 &   0.477 &   0.000 & 1\\
   \rowcolor{\ROWCOLOR}
           Eurex BUND &    CME\_CBT 30Y TB & $ -0.006 \pm   0.010$ & $ -0.000 \pm   0.003$ & $   -0.475 \pm   3.175$ &   0.609 &   0.006 &   0.280 &   0.000 & 0\\
   \rowcolor{\ROWCOLOR}
      CME\_CBT 30Y TB &    \bb{Eurex BUND} & $  0.006 \pm   0.008$ & $ -0.010 \pm   0.003$ & $   -9.862 \pm   2.550$ &   0.588 &  -0.033 &   0.330 &   0.000 & 0\\
   \bb{CME MINI S\&P} &      CME MINI NSDQ & $ -0.004 \pm   0.005$ & $  0.004 \pm   0.002$ & $    3.956 \pm   1.821$ &   0.410 &   0.024 &   0.523 &   0.039 & 0\\
   \bb{CME MINI NSDQ} &      CME MINI S\&P & $  0.021 \pm   0.004$ & $  0.008 \pm   0.002$ & $    7.532 \pm   1.783$ &   0.385 &  -0.027 &   0.529 &   0.150 & 1\\
   \rowcolor{\ROWCOLOR}
   ICE\_NYBOT MNRUS2K &      CME MINI S\&P & $  0.005 \pm   0.004$ & $ -0.002 \pm   0.002$ & $   -1.512 \pm   1.807$ &   0.379 &  -0.016 &   0.526 &   0.000 & 1\\
   \rowcolor{\ROWCOLOR}
   \bb{CME MINI S\&P} & ICE\_NYBOT MNRUS2K & $  0.022 \pm   0.005$ & $  0.016 \pm   0.002$ & $   15.427 \pm   1.814$ &   0.404 &  -0.001 &   0.519 &   0.000 & 1\\
   ICE\_NYBOT MNRUS2K &      CME MINI NSDQ & $ -0.004 \pm   0.005$ & $ -0.002 \pm   0.002$ & $   -1.685 \pm   1.967$ &   0.457 &   0.005 &   0.479 &   0.138 & 0\\
   \bb{CME MINI NSDQ} & ICE\_NYBOT MNRUS2K & $  0.037 \pm   0.005$ & $  0.017 \pm   0.002$ & $   16.092 \pm   1.948$ &   0.457 &  -0.032 &   0.457 &   0.000 & 1\\
   \rowcolor{\ROWCOLOR}
   \bb{Eurex DJEST50} &          Eurex DAX & $  0.007 \pm   0.005$ & $  0.003 \pm   0.002$ & $    2.908 \pm   2.082$ &   0.317 &  -0.008 &   0.615 &   1.000 & 1\\
   \rowcolor{\ROWCOLOR}
            Eurex DAX &      Eurex DJEST50 & $ -0.008 \pm   0.005$ & $ -0.001 \pm   0.002$ & $   -1.308 \pm   2.083$ &   0.318 &   0.019 &   0.612 &   1.000 & 1\\
     \bb{CME CMX GLD} &        CME CMX SIL & $  0.035 \pm   0.005$ & $  0.021 \pm   0.002$ & $   20.262 \pm   2.045$ &   0.437 &  -0.018 &   0.466 &   0.995 & 1\\
          CME CMX SIL &   \bb{CME CMX GLD} & $  0.002 \pm   0.005$ & $ -0.004 \pm   0.002$ & $   -4.123 \pm   2.013$ &   0.417 &  -0.017 &   0.490 &   0.013 & 0\\
   \rowcolor{\ROWCOLOR}
     \bb{CME CMX GLD} &      Forex EUR-USD & $  0.072 \pm   0.019$ & $  0.015 \pm   0.005$ & $   14.315 \pm   4.385$ &   0.778 &  -0.045 &   0.228 &   0.178 & 1\\
   \rowcolor{\ROWCOLOR}
        Forex EUR-USD &   \bb{CME CMX GLD} & $ -0.090 \pm   0.018$ & $ -0.023 \pm   0.004$ & $  -22.335 \pm   4.141$ &   0.776 &   0.050 &   0.234 &   0.998 & 1\\
     \bb{CME CMX GLD} &      CME MINI S\&P & $ -0.015 \pm   0.034$ & $  0.008 \pm   0.004$ & $    7.721 \pm   3.904$ &   0.906 &   0.021 &   0.200 &   0.000 & 0\\
        CME MINI S\&P &   \bb{CME CMX GLD} & $ -0.002 \pm   0.030$ & $ -0.018 \pm   0.004$ & $  -17.374 \pm   3.799$ &   0.896 &  -0.027 &   0.204 &   0.000 & 0\\
   \rowcolor{\ROWCOLOR}
          CME CMX GLD &          Eurex DAX & $ -0.072 \pm   0.043$ & $  0.001 \pm   0.004$ & $    1.243 \pm   4.043$ &   0.926 &   0.043 &   0.189 &   0.000 & 1\\
   \rowcolor{\ROWCOLOR}
            Eurex DAX &   \bb{CME CMX GLD} & $ -0.067 \pm   0.072$ & $ -0.016 \pm   0.006$ & $  -15.293 \pm   5.379$ &   0.944 &   0.009 &   0.165 &   0.000 & 0\\
     \bb{CME CMX GLD} &        CME PH CRDE & $  0.074 \pm   0.015$ & $  0.030 \pm   0.003$ & $   28.410 \pm   3.197$ &   0.787 &  -0.018 &   0.235 &   0.000 & 1\\
          CME PH CRDE &   \bb{CME CMX GLD} & $ -0.084 \pm   0.014$ & $ -0.040 \pm   0.003$ & $  -38.052 \pm   3.225$ &   0.762 &   0.013 &   0.242 &   0.000 & 1\\
   \rowcolor{\ROWCOLOR}
          CME PH CRDE &     \bb{Eurex DAX} & $ -0.128 \pm   0.018$ & $ -0.042 \pm   0.004$ & $  -40.272 \pm   3.367$ &   0.823 &   0.040 &   0.225 &   0.147 & 1\\
   \rowcolor{\ROWCOLOR}
            Eurex DAX &        CME PH CRDE & $  0.038 \pm   0.022$ & $  0.002 \pm   0.005$ & $    2.370 \pm   4.313$ &   0.830 &  -0.027 &   0.172 &   0.000 & 1\\
          CME PH CRDE & \bb{Forex EUR-USD} & $ -0.157 \pm   0.018$ & $ -0.074 \pm   0.005$ & $  -70.513 \pm   4.453$ &   0.765 &   0.021 &   0.216 &   1.000 & 1\\
   \bb{Forex EUR-USD} &        CME PH CRDE & $  0.179 \pm   0.019$ & $  0.074 \pm   0.004$ & $   70.949 \pm   4.291$ &   0.789 &  -0.039 &   0.219 &   0.989 & 1\\
   \rowcolor{\ROWCOLOR}
          CME PH CRDE &     \bb{CME PH NG} & $ -0.171 \pm   0.027$ & $ -0.035 \pm   0.004$ & $  -33.655 \pm   3.786$ &   0.882 &   0.059 &   0.180 &   1.000 & 1\\
   \rowcolor{\ROWCOLOR}
       \bb{CME PH NG} &        CME PH CRDE & $  0.204 \pm   0.024$ & $  0.041 \pm   0.004$ & $   39.363 \pm   3.698$ &   0.871 &  -0.078 &   0.183 &   0.737 & 1\\
   \bottomrule
   \end{tabular}
  }
  \caption{Results on 60\,\textrm{min} chart (time of extrema confirmed). $^*$This market leads the other one.}
  \label{tab:60min_confirmed}
\end{sidewaystable}

First we discuss the results for the time of extrema and afterwards the results for the confirmation time of the extrema.

\paragraph{Time of extrema}
First we note that the results are mostly independent of the mean wavelength which we can see from the additional information of each bin, i.e. the minimal and maximal value for this bin and the standard deviation. Next we see a very weak correlation between EUR-USD vs. JPY-USD, Gold vs. EUR-USD, Gold vs. S\&P 500, Gold vs. FDAX, Gold vs. Oil, Oil vs. FDAX and Oil vs. EUR-USD. The pairs of markets also have a relatively large standard deviation $\hat{S}$ and small concentration around its mean indicated by the small kurtosis $\hat{k}$.

All other combinations of markets illustrated in Table~\ref{tab:60min} and Figures~\ref{fig:res1} and \ref{fig:res3} to \ref{fig:res10} show a large peak near the mean angular direction between \SI{20}{\percent} up to \SI{53}{\percent}. This means that the probability is significantly high that extreme values for both markets are shaped in almost the exact time. Of course this leads to smaller standard deviations and higher kurtosis.

\paragraph{Confirmation time of extrema}
Since the point in time of confirming an extreme value by the MinMax process is more sensitive to the price development than the very fixed point in time of the extreme value itself we already expect scattered observations. However even here we can see a peak in the mean angular direction of about half of the size of the peak for the time of extrema of the strongly correlated pairs of markets. The values in Table~\ref{tab:60min_confirmed} are approximately of the same order as in Table~\ref{tab:60min}.

\paragraph{All together} We see strong correlations for extrema and confirmed extrema between combinations of FDAX, Euro-Bund, Euro STOXX, S\&P 500, U.S. Treasury, NASDAQ 100, Russel 2000 and between the foreign exchanges except EUR-USD versus JPY-USD. Additionally Gold and Silver has a strong correlation whereas all other combinations with at least one market from commodities seem to be weakly correlated or even nearly uncorrelated. Thus from the point of view of local extreme values the commodities are separated from other markets.

The lead-lag $\ell^{(w)}$, see Section~\ref{sec:3.3}, is between \SI{5}{\minute} and \SI{10}{\minute} for the point in time of the extrema for the indexes and foreign exchanges and also for Gold versus Silver. Note that this is just a fraction of the duration of one single period of the \SI{60}{\minute} chart. Even the points in time of the extrema are just the time stamp of a candle and not the exact time of the extreme value itself, i.e. these points in time have an uncertainty of $\SI{+-30}{\minute}$. Therefore we cannot view the value $\ell^{(w)}$ as an absolute value but more as a tendency of the lead or lag for the candles in which the extreme values occur.

\begin{remark}
  In most of the cases our investigations of the correlation of two markets yields one market leading and one market following, e.g. DAX Futures leads E-mini S\&P 500 Futures, no matter which one is considered primary or secondary market. Note however, that in some cases the leading market is not unique like for instance the Gold Futures versus Silver Futures, or, sometimes our calculation cannot decide which market is leading.
\end{remark}

\def\FIGWIDTH{0.9\linewidth}

\begin{figure}
  \centering
  \includegraphics[width=\FIGWIDTH]{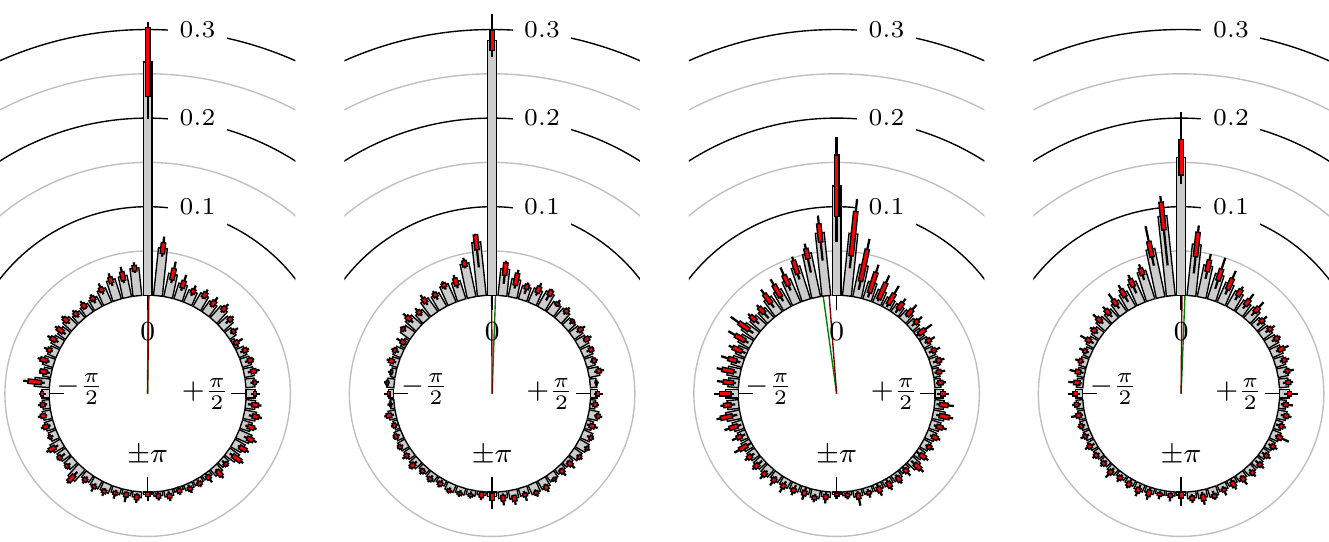}
  \caption{DAX Futures versus E-mini S\&P 500 Futures}
  \label{fig:res1}
\end{figure}

\begin{figure}
  \centering
  \includegraphics[width=\FIGWIDTH]{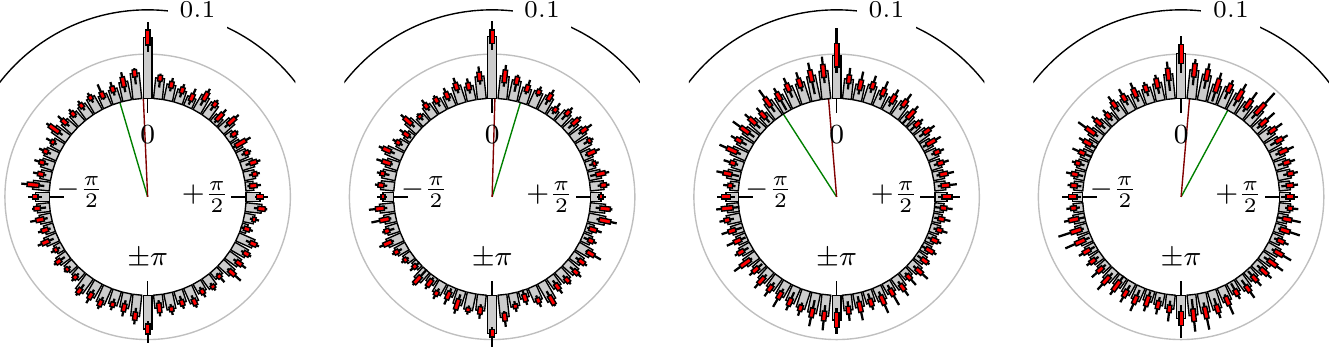}
  \caption{EUR-USD versus JPY-USD}
\end{figure}

\begin{figure}
  \centering
  \includegraphics[width=\FIGWIDTH]{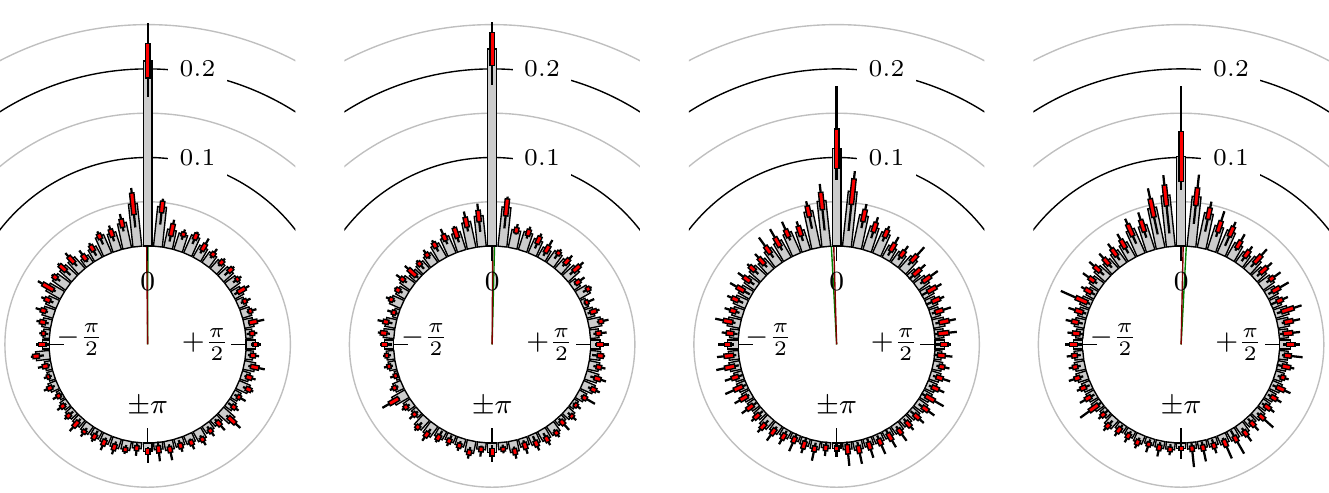}
  \caption{EUR-USD versus GBP-USD}
  \label{fig:res3}
\end{figure}

\begin{figure}
  \centering
  \includegraphics[width=\FIGWIDTH]{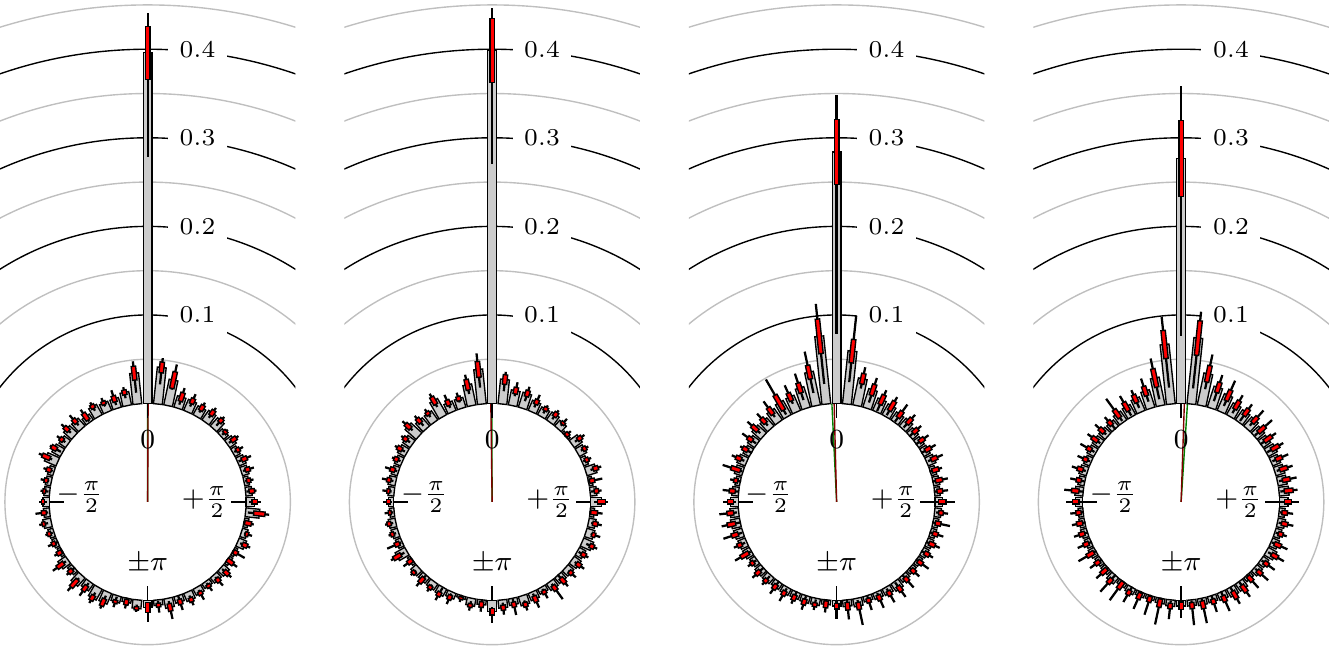}
  \caption{EUR-USD versus CHF-USD}
  \label{fig:res4}
\end{figure}

\begin{figure}
  \centering
  \includegraphics[width=\FIGWIDTH]{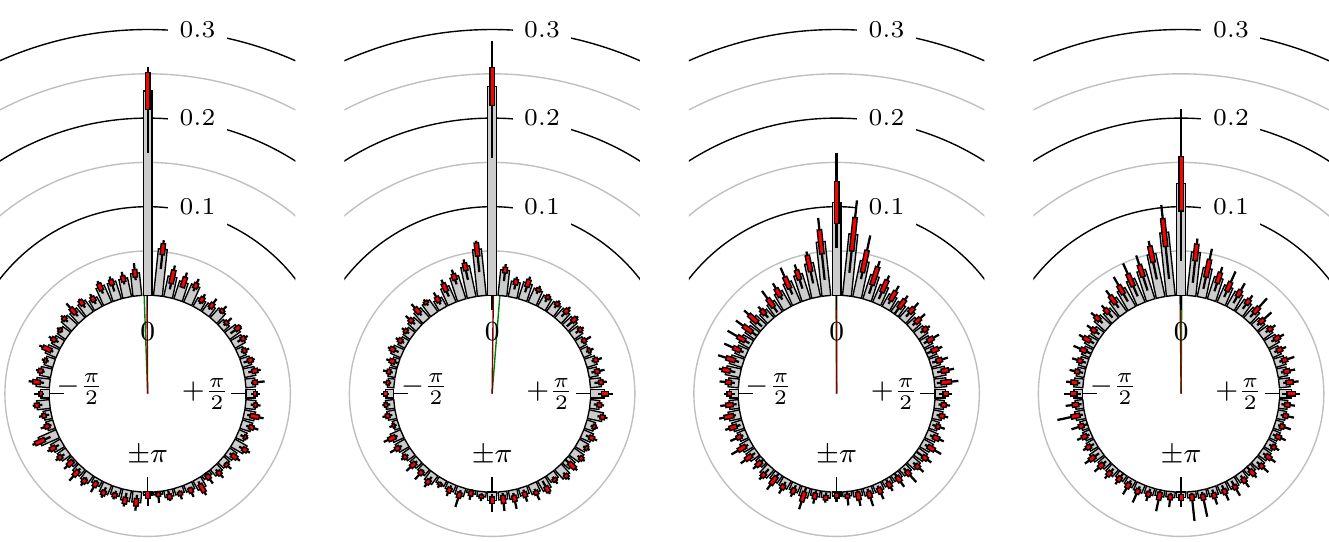}
  \caption{Euro-Bund Futures versus U.S. Treasury Bond Futures}
\end{figure}

\begin{figure}
  \centering
  \includegraphics[width=\FIGWIDTH]{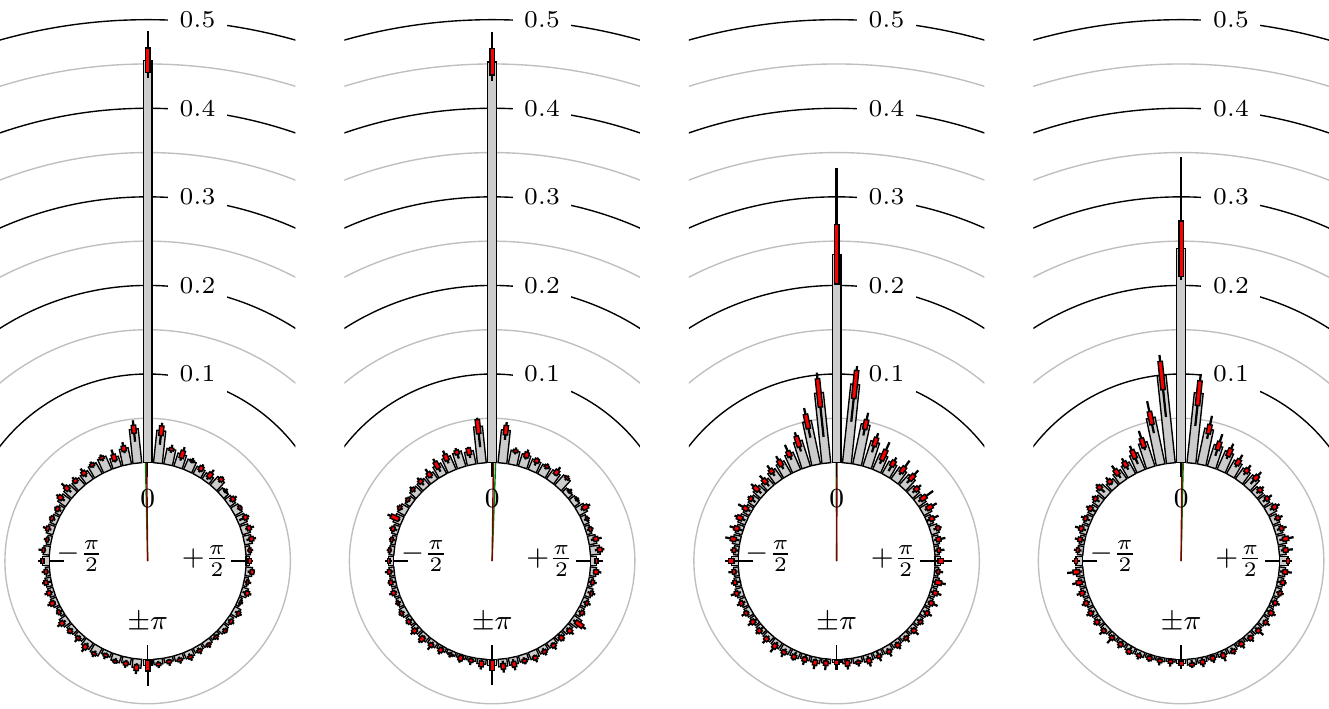}
  \caption{E-mini S\&P 500 Futures versus E-mini NASDAQ 100 Futures}
\end{figure}

\begin{figure}
  \centering
  \includegraphics[width=\FIGWIDTH]{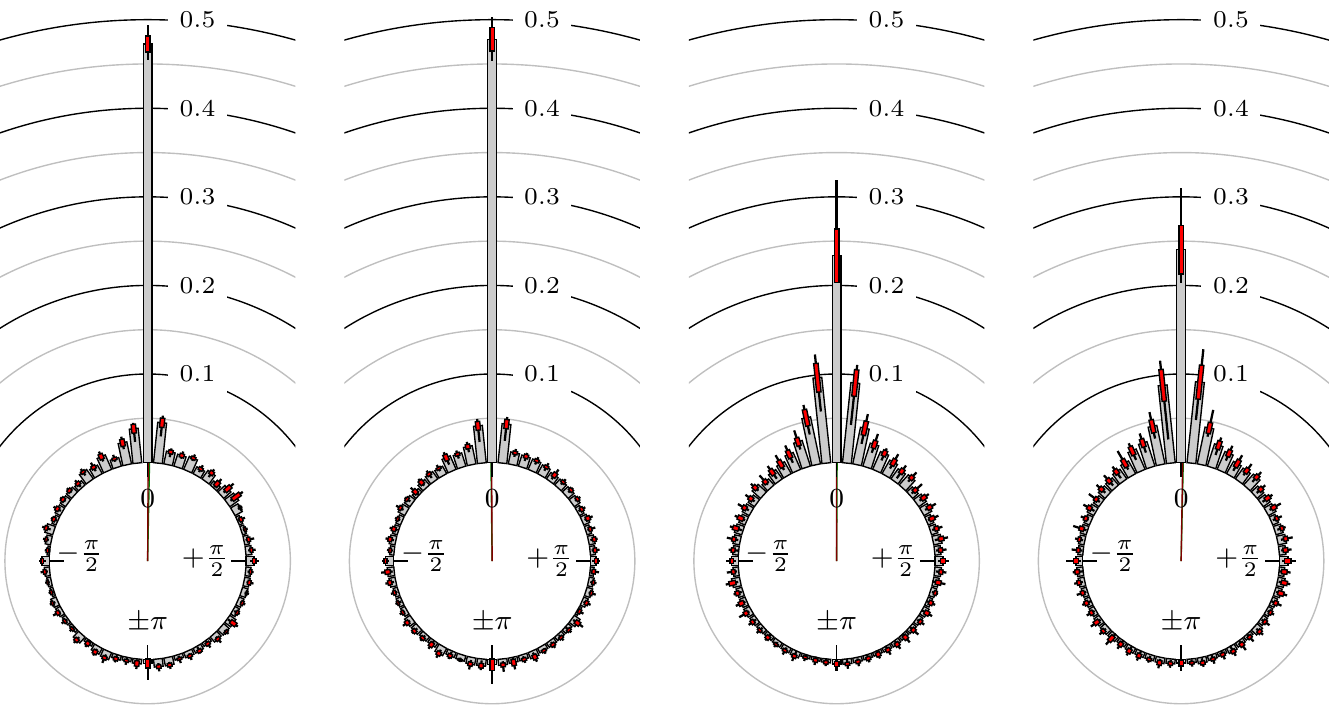}
  \caption{Russell 2000 Index Mini Futures versus E-mini S\&P 500 Futures}
\end{figure}

\begin{figure}
  \centering
  \includegraphics[width=\FIGWIDTH]{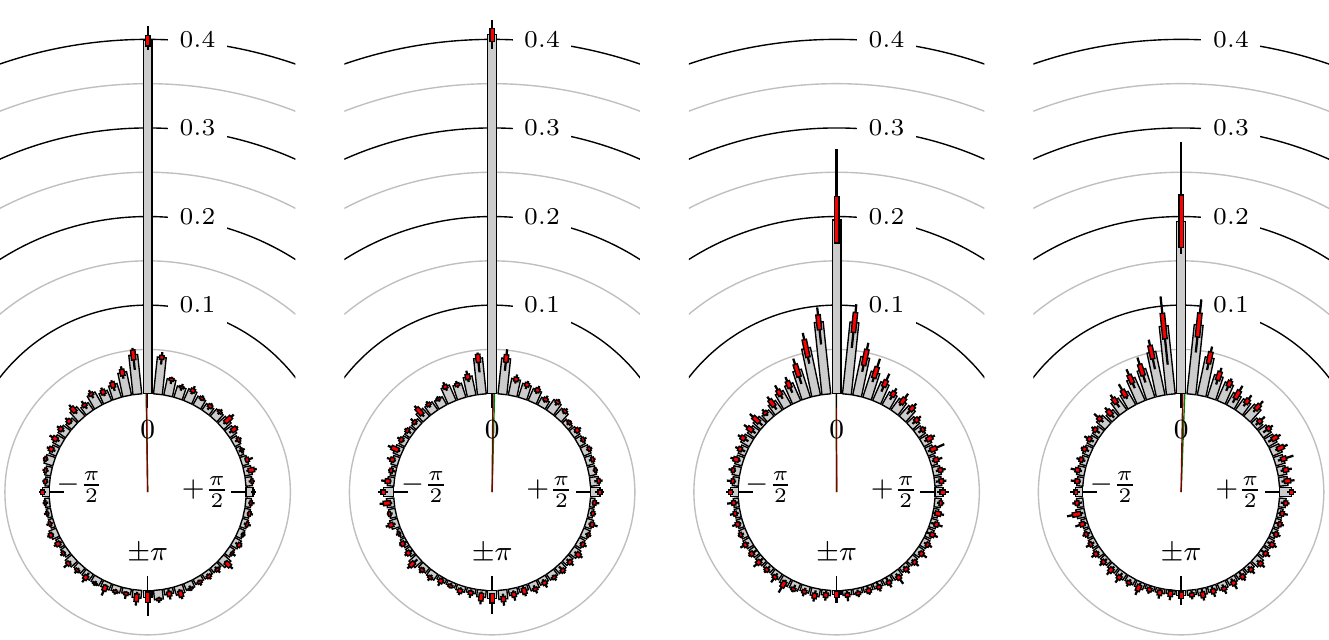}
  \caption{Russell 2000 Index Mini Futures versus E-mini NASDAQ 100 Futures}
\end{figure}

\begin{figure}
  \centering
  \includegraphics[width=\FIGWIDTH]{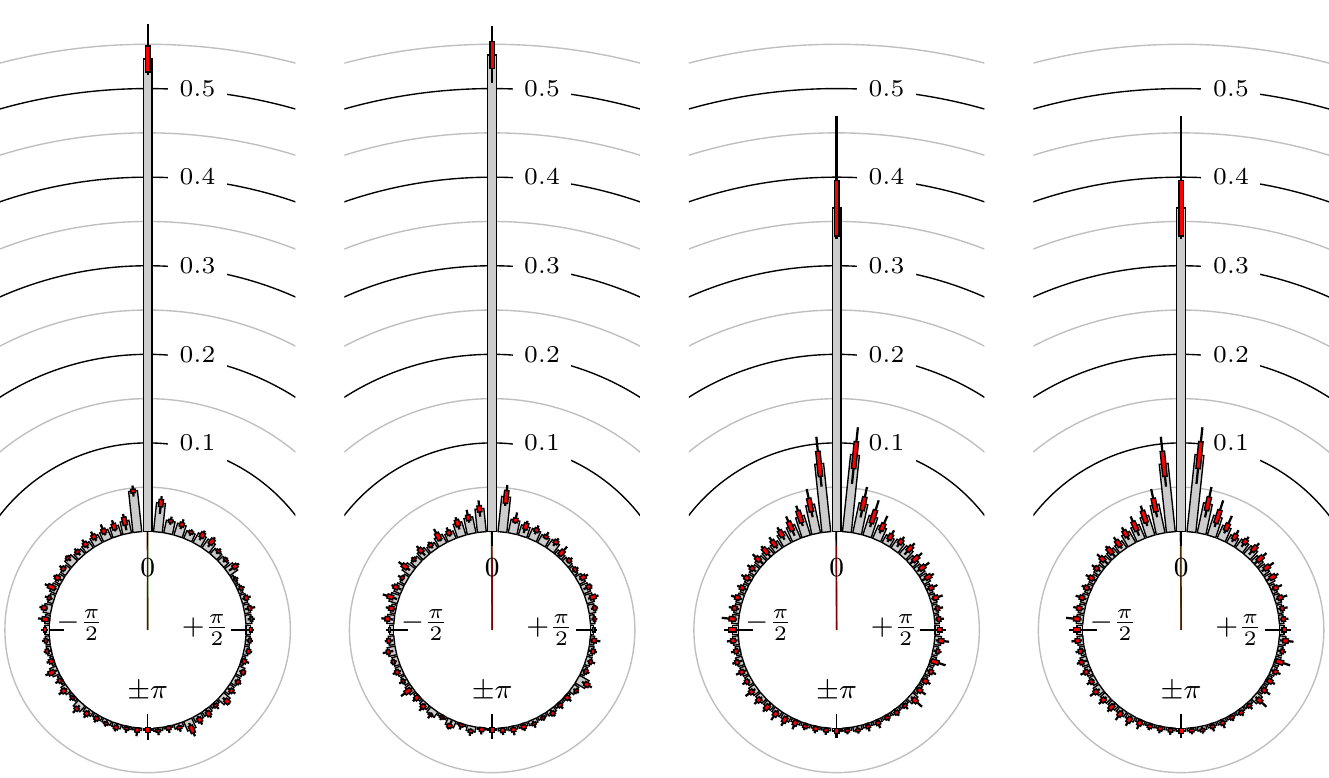}
  \caption{EURO STOXX 50 Index Futures versus DAX Futures}
\end{figure}

\begin{figure}
  \centering
  \includegraphics[width=\FIGWIDTH]{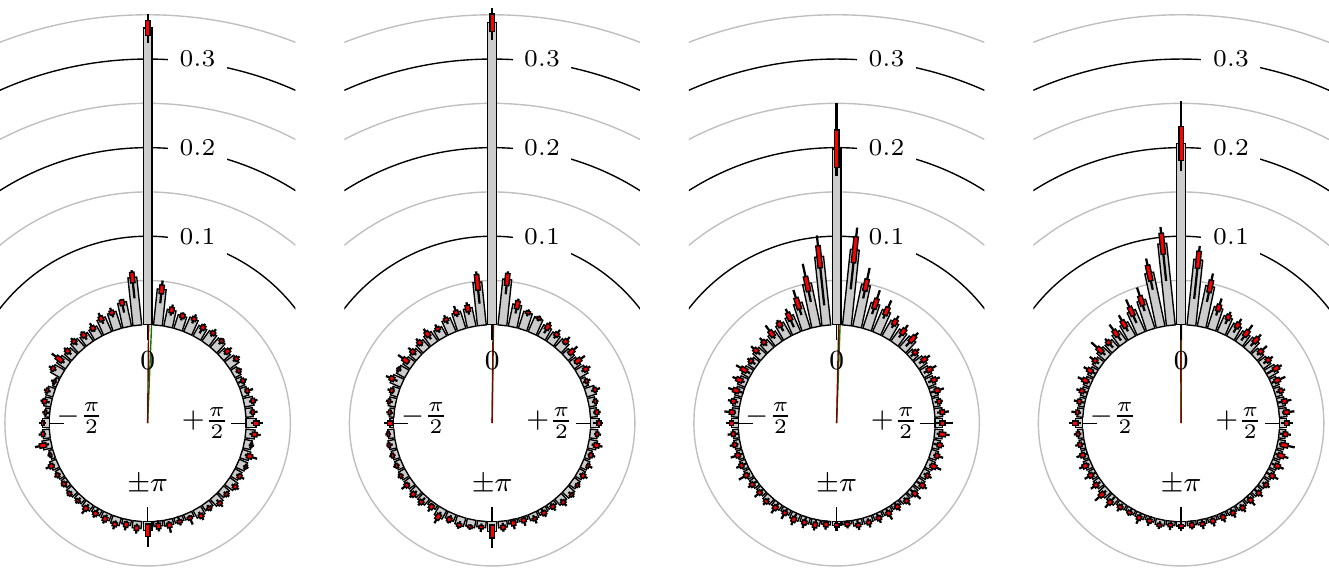}
  \caption{Gold Futures versus Silver Futures}
  \label{fig:res10}
\end{figure}

\begin{figure}
  \centering
  \includegraphics[width=\FIGWIDTH]{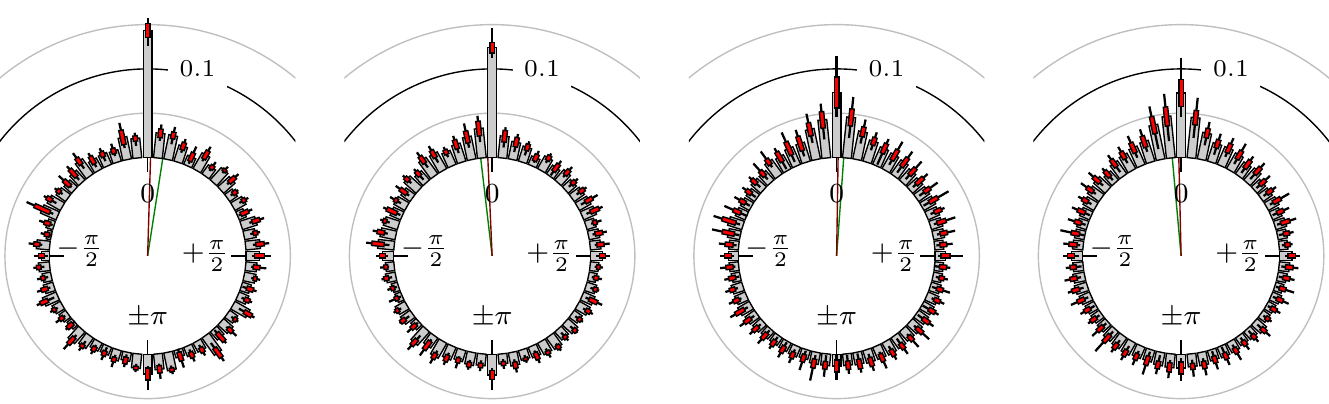}
  \caption{Gold Futures versus EUR-USD}
\end{figure}

\begin{figure}
  \centering
  \includegraphics[width=\FIGWIDTH]{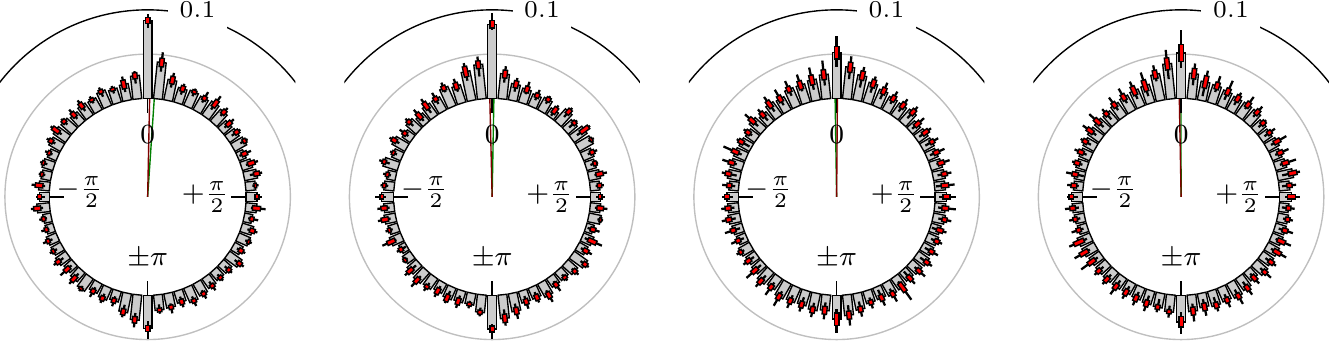}
  \caption{Gold Futures versus E-mini S\&P 500 Futures}
\end{figure}

\begin{figure}
  \centering
  \includegraphics[width=\FIGWIDTH]{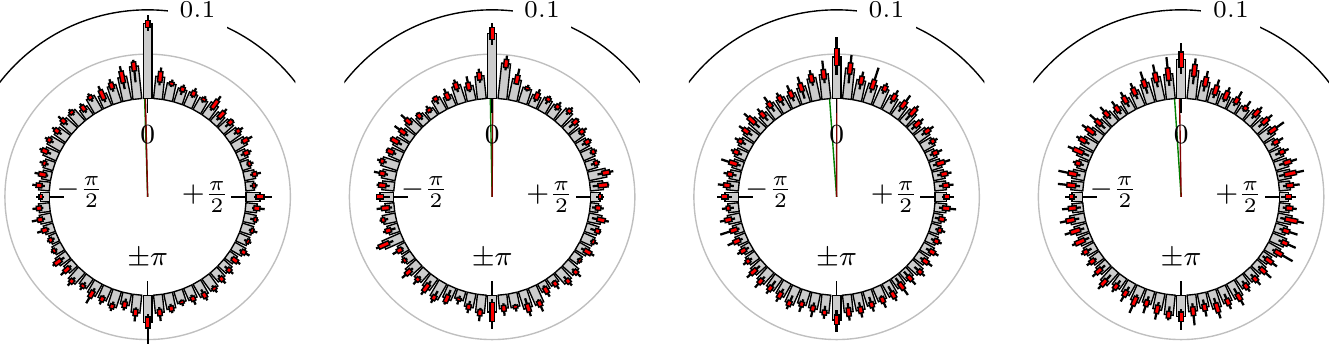}
  \caption{Gold Futures versus DAX Futures}
\end{figure}

\begin{figure}
  \centering
  \includegraphics[width=\FIGWIDTH]{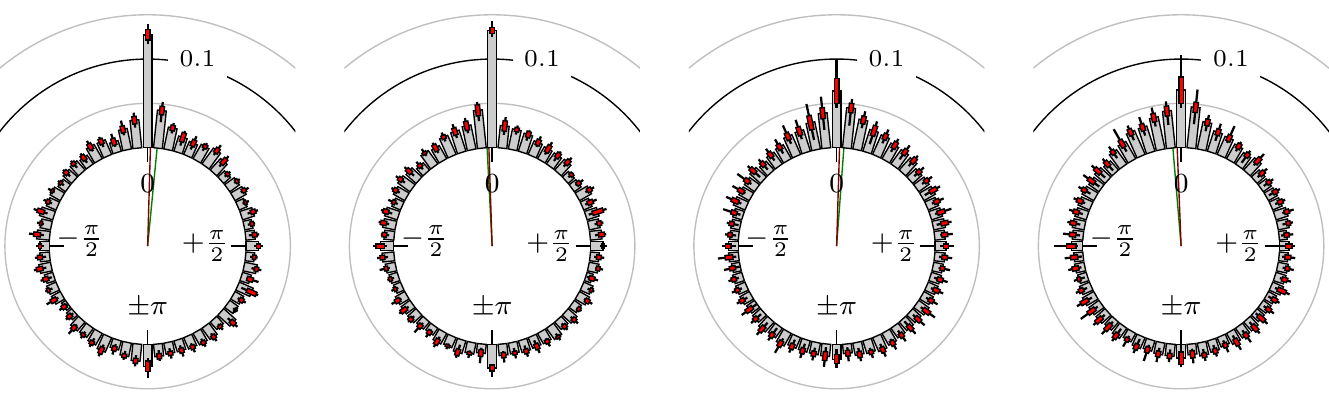}
  \caption{Gold Futures versus Crude Oil Futures}
\end{figure}

\begin{figure}
  \centering
  \includegraphics[width=\FIGWIDTH]{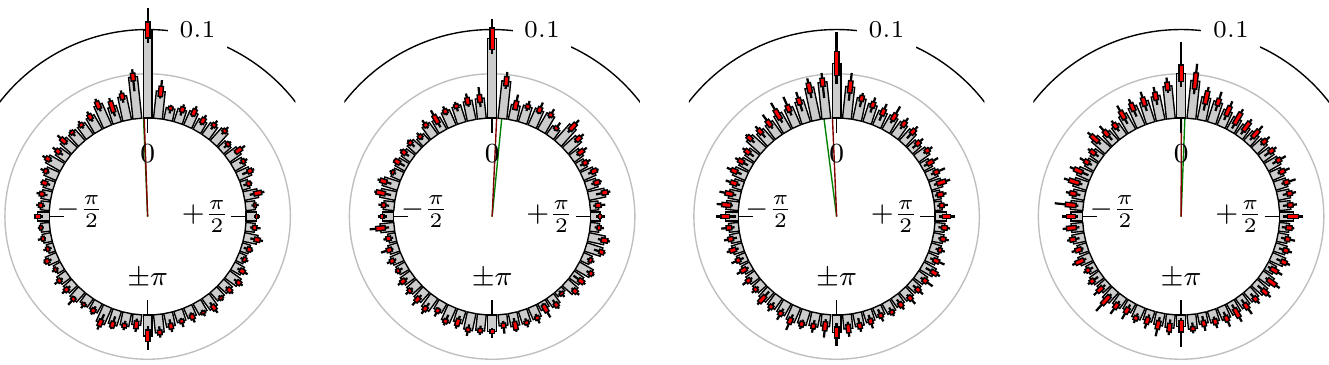}
  \caption{Crude Oil Futures versus DAX Futures}
\end{figure}

\begin{figure}
  \centering
  \includegraphics[width=\FIGWIDTH]{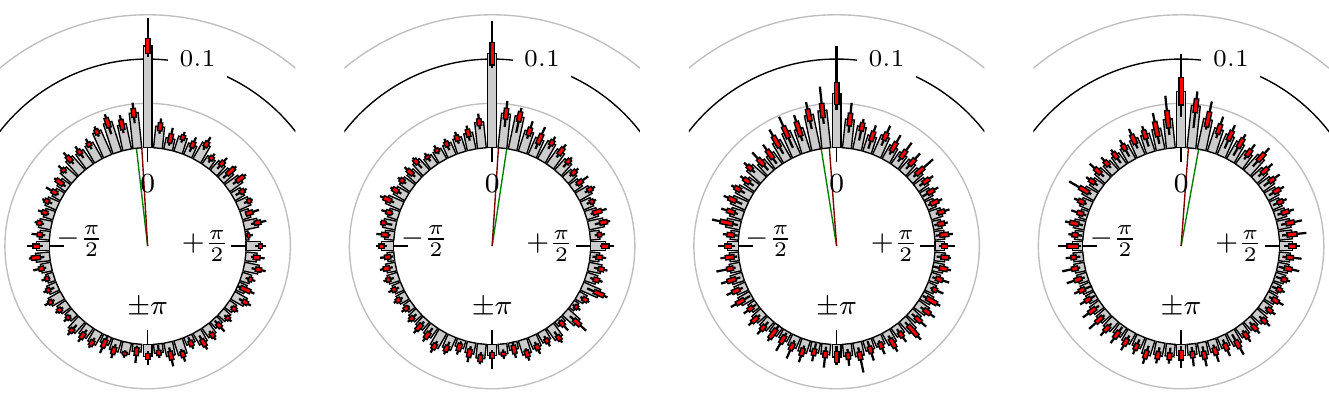}
  \caption{Crude Oil Futures versus EUR-USD}
\end{figure}

\begin{figure}
  \centering
  \includegraphics[width=\FIGWIDTH]{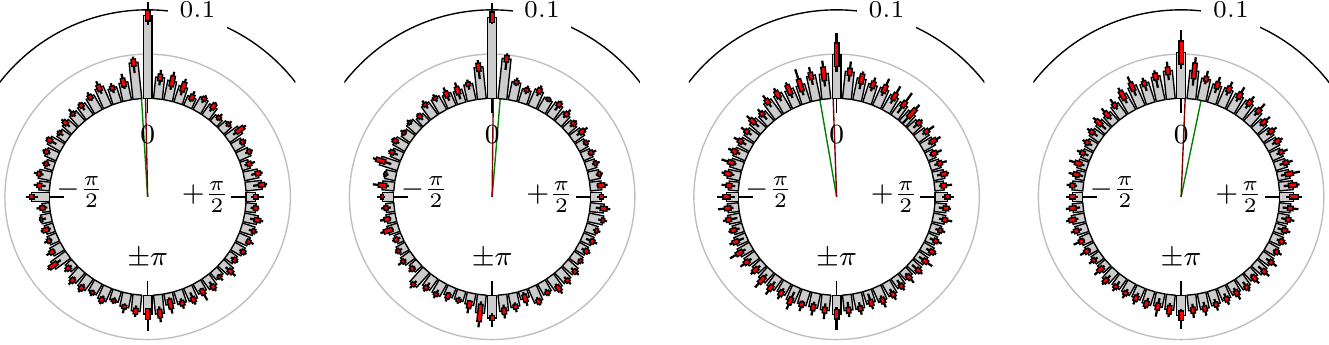}
  \caption{Crude Oil Futures versus Natural Gas (Henry Hub) Physical Futures}
  \label{fig:res17}
\end{figure}

\begin{figure}
  \centering
  \includegraphics[width=\FIGWIDTH]{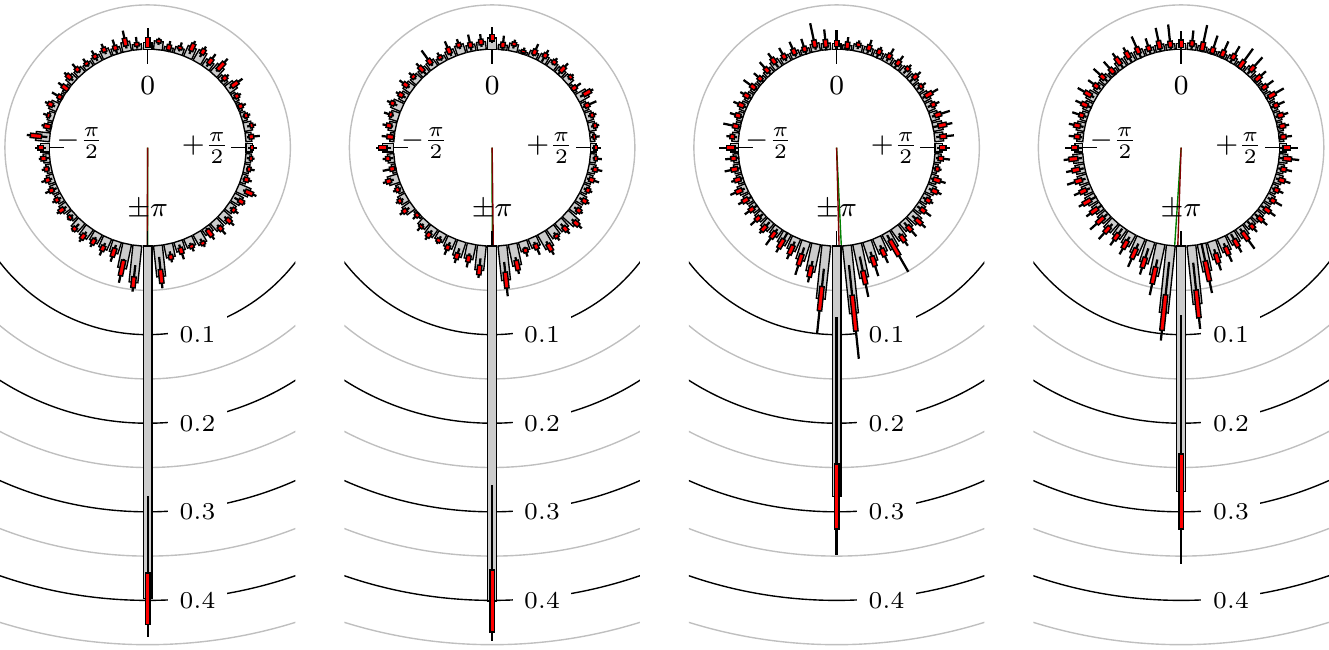}
  \caption{EUR-USD versus USD-CHF (cf. Figure~\ref{fig:res4})}
  \label{fig:res_negative_corr}
\end{figure}

\begin{remark}
  For the currency Swiss franc it is more common to analyze USD-CHF instead of CHF-USD as we do in the above discussion. The reason we focus on CHF-USD is to see the positive correlation to EUR-USD and thus to have a more natural interpretation for lead and lag as in Definition~\ref{def:lead_lag}.

  However, it is also possible to compare (strongly) negative correlated markets as EUR-USD versus USD-CHF. In Figure~\ref{fig:res_negative_corr} we see the results for this combination. We expect that the results are the same as for the combination EUR-USD versus CHF-USD but shifted by $\pi$. If we compare Figures~\ref{fig:res4} and \ref{fig:res_negative_corr} we actually see this connection perfectly. This is also the case for the Japanese yen.
\end{remark}

\section{Conclusion and outlook}\label{sec:5}

We introduced the notion of lead-lag relationship from a market technical point of view. Using the local extreme values of the markets we get an empirical distribution of their phase shifts on the unit sphere. The directional statistics helps us to illustrate and quantify the results.

We observed many strongly correlated pairs of markets with respect to their extreme values while, of course, there are combinations with a very weak connection. Combinations of indexes show the highest correlation and also a measurable lead or lag. Since we use a geometrical approach based on the actual local extreme values of the chart, i.e. on some kind of reversal points, the results can directly be used for trading strategies.

In future work the authors plan to localize this method to shorter time intervals so that we obtain even more meaningful results for live/real time data.


\end{document}